\documentclass[12pt,epsf]{article}
\usepackage{graphicx,amsmath,amssymb}
\begin{document}

\def\a{\alpha}
\def\b{\beta}
\def\c{\varepsilon}
\def\d{\delta}
\def\e{\epsilon}
\def\f{\phi}
\def\g{\gamma}
\def\h{\theta}
\def\k{\kappa}
\def\l{\lambda}
\def\m{\mu}
\def\n{\nu}
\def\p{\psi}
\def\q{\partial}
\def\r{\rho}
\def\s{\sigma}
\def\t{\tau}
\def\u{\upsilon}
\def\v{\varphi}
\def\w{\omega}
\def\x{\xi}
\def\y{\eta}
\def\z{\zeta}
\def\D{\Delta}
\def\G{\Gamma}
\def\H{\Theta}
\def\L{\Lambda}
\def\F{\Phi}
\def\P{\Psi}
\def\S{\Sigma}

\def\o{\over}
\def\beq{\begin{eqnarray}}
\def\eeq{\end{eqnarray}}
\newcommand{\gsim}{ \mathop{}_{\textstyle \sim}^{\textstyle >} }
\newcommand{\lsim}{ \mathop{}_{\textstyle \sim}^{\textstyle <} }
\newcommand{\vev}[1]{ \left\langle {#1} \right\rangle }
\newcommand{\bra}[1]{ \langle {#1} | }
\newcommand{\ket}[1]{ | {#1} \rangle }
\newcommand{\EV}{ {\rm eV} }
\newcommand{\KEV}{ {\rm keV} }
\newcommand{\MEV}{ {\rm MeV} }
\newcommand{\GEV}{ {\rm GeV} }
\newcommand{\TEV}{ {\rm TeV} }
\def\diag{\mathop{\rm diag}\nolimits}
\def\Spin{\mathop{\rm Spin}}
\def\SO{\mathop{\rm SO}}
\def\O{\mathop{\rm O}}
\def\SU{\mathop{\rm SU}}
\def\U{\mathop{\rm U}}
\def\Sp{\mathop{\rm Sp}}
\def\SL{\mathop{\rm SL}}
\def\tr{\mathop{\rm tr}}

\def\IJMP{Int.~J.~Mod.~Phys. }
\def\MPL{Mod.~Phys.~Lett. }
\def\NP{Nucl.~Phys. }
\def\PL{Phys.~Lett. }
\def\PR{Phys.~Rev. }
\def\PRL{Phys.~Rev.~Lett. }
\def\PTP{Prog.~Theor.~Phys. }
\def\ZP{Z.~Phys. }

\newcommand{\bea}{\begin{eqnarray}}   
\newcommand{\eea}{\end{eqnarray}}
\newcommand{\bear}{\begin{array}}  
\newcommand {\eear}{\end{array}}
\newcommand{\bef}{\begin{figure}}  
\newcommand {\eef}{\end{figure}}
\newcommand{\bec}{\begin{center}}  
\newcommand {\eec}{\end{center}}
\newcommand{\non}{\nonumber}  
\newcommand {\eqn}[1]{\beq {#1}\eeq}
\newcommand{\la}{\left\langle}  
\newcommand{\ra}{\right\rangle}
\newcommand{\ds}{\displaystyle}
\def\SEC#1{Sec.~\ref{#1}}
\def\FIG#1{Fig.~\ref{#1}}
\def\EQ#1{Eq.~(\ref{#1})}
\def\EQS#1{Eqs.~(\ref{#1})}
\def\TEV#1{10^{#1}{\rm\,TeV}}
\def\GEV#1{10^{#1}{\rm\,GeV}}
\def\MEV#1{10^{#1}{\rm\,MeV}}
\def\KEV#1{10^{#1}{\rm\,keV}}
\def\lrf#1#2{ \left(\frac{#1}{#2}\right)}
\def\lrfp#1#2#3{ \left(\frac{#1}{#2} \right)^{#3}}
\def\REF#1{Ref.~\cite{#1}}
\newcommand{\mc}{m_\chi}
\newcommand{\mph}{m_\phi}


\baselineskip 0.7cm

\begin{titlepage}

\begin{flushright}
TU-910\\
IPMU12-0114\\
UT-12-13
\end{flushright}

\vskip 1.35cm
\begin{center}
{\large \bf 
Alchemical Inflation: inflaton turns into Higgs
}
\vskip 1.2cm
Kazunori Nakayama$^{a,c}$
and
Fuminobu Takahashi$^{b,c}$

\vskip 0.4cm

{\it $^a$Department of Physics, University of Tokyo, Tokyo 113-0033, Japan}\\
{\it $^b$Department of Physics, Tohoku University, Sendai 980-8578, Japan}\\
{\it $^c$ Kavli Institute for the Physics and Mathematics of the Universe,
University of Tokyo,  Kashiwa, 277-8583, Japan}\\

\vskip 1.5cm

\abstract{ We propose a new inflation model in which a gauge singlet
  inflaton turns into the Higgs condensate after inflation. The
  inflationary path is characterized by a moduli space of
  supersymmetric vacua spanned by the inflaton and Higgs field.  The
  inflation energy scale is  related to the soft supersymmetry
  breaking, and the Hubble parameter during inflation is smaller than
  the gravitino mass. The initial condition for the successful
  inflation is naturally realized by the pre-inflation in which the
  Higgs plays a role of the waterfall field.  }
\end{center}
\end{titlepage}

\setcounter{page}{2}

\section{Introduction}

The existence of the inflationary era~\cite{Guth:1980zm} in the early
Universe is strongly suggested by the
observations~\cite{Komatsu:2010fb}. In particular, the density
perturbations extending beyond the Hubble horizon at the last
scattering surface can be interpreted as the evidence for the
accelerated expansion of the Universe in the past.

While there are many inflation models, we still do not know which
inflation model is realized in nature.  The study of density
perturbations such as isocurvature perturbations, non-gaussianity,
tensor-mode, and their effects on the CMB power spectrum is a powerful
diagnostic of the mechanism that laid down the primordial density
fluctuations, but it is not enough at present to pin down the
inflation model.  This is partly because of our ignorance of thermal
history of the Universe beyond the standard big bang theory,
especially, how the inflaton reheated the Universe.

In the reheating process the inflaton transfers its energy to the
visible sector. Usually it is assumed that it proceeds through 
perturbative or non-perturbative decay of the inflaton. After
reheating, the visible sector is thermalized sooner or later, leading
to the hot radiation dominated Universe.

The reheating process is also useful to constrain theory beyond the
standard model (SM) because unwanted relics such as
gravitinos~\cite{Weinberg:zq,Krauss:1983ik} are efficiently produced
at the reheating.  Recently there was progress in this regard; the
inflaton with a finite vacuum expectation value (VEV) is necessarily
coupled to the SM sector at both tree and one-loop
levels~\cite{Endo:2006qk,Endo:2007ih}, and so, the inflaton naturally
decays and reheats the SM sector.  However it turned out that such
inflaton is also coupled to the supersymmetry (SUSY) breaking sector,
and the inflaton decay generically produces gravitinos as
well~\cite{Kawasaki:2006gs,Asaka:2006bv,Endo:2007ih,Endo:2007sz}. The
abundance of the non-thermally produced gravitinos is inversely
proportional to the reheating temperature, and so, it tightly
constrains the inflation models, especially if combined with the
thermal production of the gravitinos.

\vspace{5mm}

In this letter, we propose a new inflation model, in which a SM gauge
singlet inflaton turns into the Higgs field (or more precisely, $H_u
H_d$) after inflation. The energy stored in the inflaton is directly
transferred to the $H_u H_d$ condensate, where $H_u$ and $H_d$ are up-
and down-type Higgs fields, respectively.  This is a new type of
reheating; instead of producing the SM particles from the inflaton
decay, the inflaton energy is directly transmuted into the $H_u H_d$
flat direction. As we will see shortly, the transmutation is due to
the inflaton dynamics in the moduli space of supersymmetric vacua
spanned by the inflaton and the Higgs field.  Interestingly, because
of this transmutation, the new inflation ends up in a symmetry vacuum
of the Higgs boson, which enables rapid thermalization of the Higgs
condensate.  The inflation model has the following several
cosmological and phenomenological virtues.

\begin{itemize}

\item The inflaton potential is induced by the SUSY breaking, and so,
  the SUSY breaking scale is directly related to the normalization of
  the density perturbations.

\item Non-thermal gravitino production does not occur, both because
  the Higgs flat direction has a small VEV and because the soft mass
  for $H_u H_d$ is comparable to the gravitino mass.

\item The inflation scale is lower than the gravitino mass by more
  than one order of magnitude:
\beq
H_{\rm inf} \ll m_{3/2}.
\eeq
 This is advantageous from the point of view of the moduli
 stabilization~\cite{Kallosh:2004yh}.

\item The initial condition of the inflation can be naturally realized
  by the primordial hybrid inflation~\cite{Copeland:1994vg} or its
  variant, the smooth hybrid inflation~\cite{Lazarides:1995vr}.

\end{itemize}

It is also possible to replace the role of the $H_u H_d$ with other
D-flat directions of the SUSY SM (SSM). Then, the Affleck-Dine (AD)
mechanism~\cite{Affleck:1984fy,Dine:1995uk} naturally works. What is
interesting is that the AD field plays an important role in the
inflationary dynamics, and its initial large VEV is a requisite for
the inflation to take place.

Our model has similarity with the so-called MSSM
inflation~\cite{Allahverdi:2006iq} in that the inflaton potential is
induced by the SUSY breaking and the energy density after inflation is
dominated by coherent oscillations of a D-flat direction in the SSM.
However, there are great differences: (1) the inflation is mainly
driven by a gauge singlet field, not the MSSM flat direction; (2) the
energy transfer of the inflaton to the visible sector is due to
non-trivial dynamics in the moduli space; (3) the AD mechanism works;
(4) the initial condition can be dynamically set by the pre-inflation.

\vspace{5mm}

The rest of this letter is organized as follows. In Sec.~\ref{sec:2}
we give the inflation model and study its dynamics in detail. The
various implications of the model are discussed in Sec.~\ref{sec:3}.
The last section is devoted for conclusions.

\section{Inflation Model}
\label{sec:2}
We consider the following superpotential:
\beq
W\;=\; S(\mu^2 - \lambda \chi^m - g  \phi^n),
\label{W}
\eeq
where $S$, $\chi$ and $\phi$ are chiral superfields, $m$ and $n$ are
integers, and $\mu$, $\lambda$ and $g$ are taken to be real and
positive by phase redefinition of the fields without loss of
generality.  We adopt the Planck units, in which $M_p = 2.4 \times
\GEV{18} $ is set to be unity.  The above form of the superpotential
can be ensured by assigning R-charges as $R(S) = 2$, and $R(\chi^m) =
R(\phi^n) = 0$ and appropriate discrete symmetry on $\phi$ and $\chi$.
Later we will identify $\chi^2 = H_u H_d$ or other flat directions.

The SUSY vacua is characterized  by $\chi$ and $\phi$ satisfying
\beq
 \lambda \chi^m + g \phi^n = \mu^2.
\label{susy}
\eeq
The direction orthogonal to the moduli space has a large SUSY mass. In
the low energy, we can integrate out those heavy degrees of freedom
and focus on the dynamics of the moduli.  There are two special
symmetry-enhanced points in the moduli space, i.e., $\chi=0$ and $\phi
= 0$. As we will see shortly, the new inflation~\cite{Linde:1981mu}
takes place as the field moves from one to the other.

Let us now introduce SUSY breaking which lifts the degeneracy of the moduli space.
Assuming the gravity mediation for the scalar mass,
we obtain
\beq
V_{\rm soft}\;=\; \mc^2 |\chi|^2 + \mph^2 |\phi|^2,
\eeq
which is considered to be valid up to the Planck scale.  The typical
scale of $\mc$ is given by the gravitino mass, $m_{3/2}$, while $\mph$
must be suppressed for successful inflation as we will see later.  We
assume that the soft SUSY breaking scale is much lower than the SUSY
mass scale in (\ref{W}), namely,
\beq
\mc \sim m_{3/2} \;\ll\; {\rm Min}\left[ m \lambda \lrfp{\mu^2}{\lambda}{\frac{m-1}{m}}, ~n g \lrfp{\mu^2}{g}{\frac{n-1}{n}} \right].
\eeq
Then the scalar potential is obtained by substituting the SUSY
condition (\ref{susy}) into $V_{\rm soft}$.

\begin{figure}[t]
\begin{center}
\includegraphics[scale=0.35]{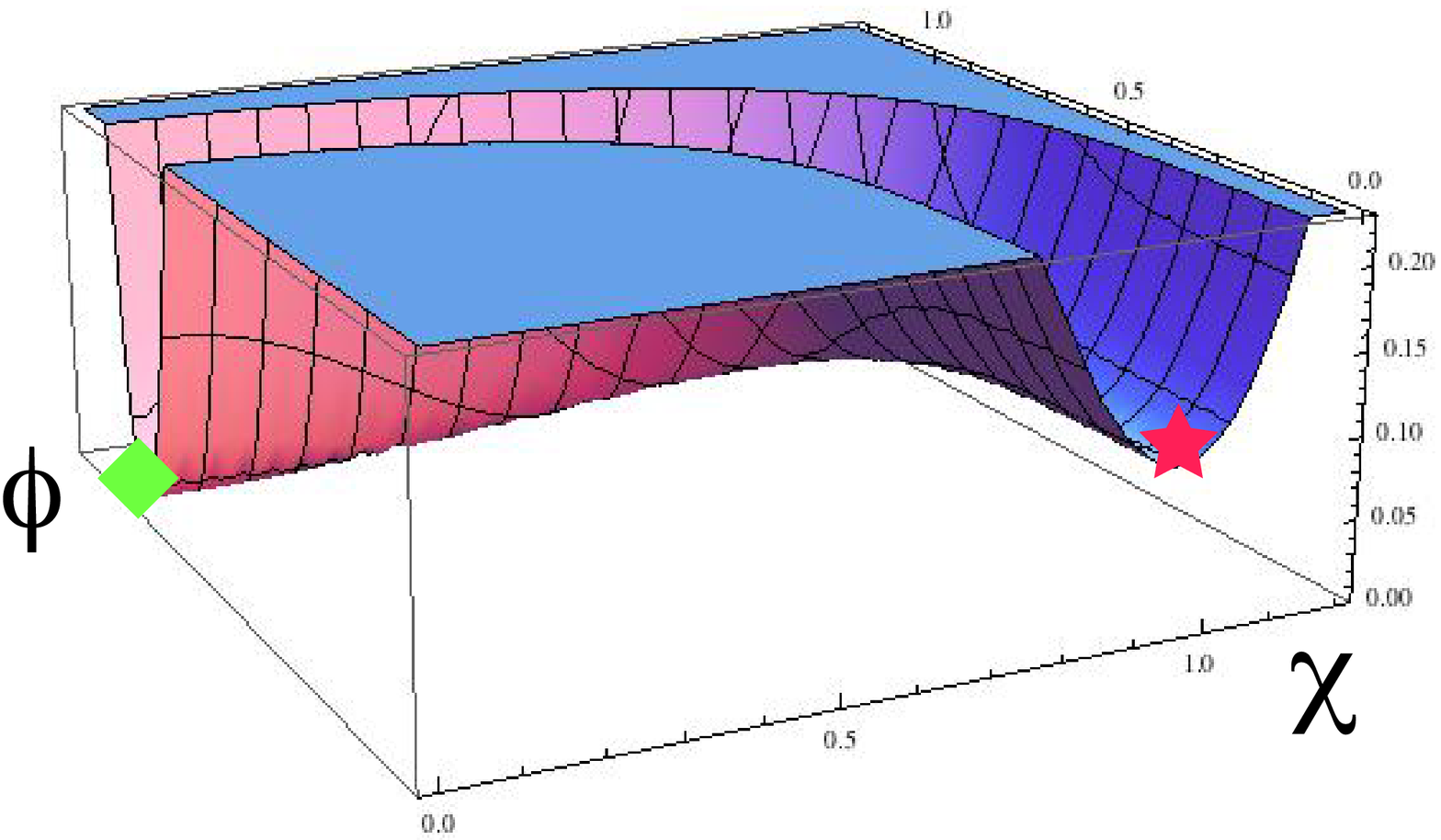}\qquad
\includegraphics[scale=0.55]{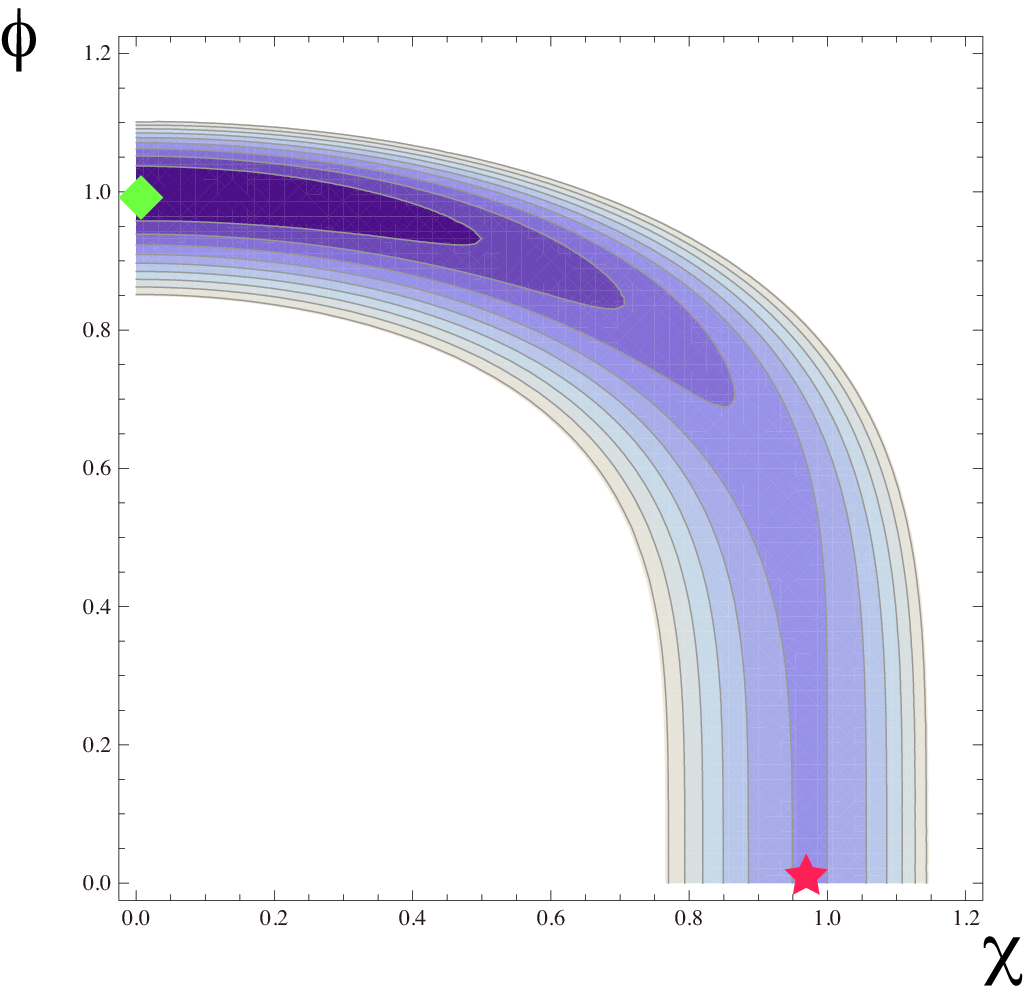}
\caption{The shape of the inflaton potential. The right panel shows the contours of the inflaton potential in the left panel.  
The star and diamond denote the potential maximum and minimum in the moduli space, respectively. The inflation takes
place along the valley of the potential near the maximum. 
}
\label{fig:potential}
\end{center}
\end{figure}

First let us study the scalar potential in the case of $m_\phi^2 =
0$. The moduli space is then lifted by the soft SUSY breaking mass of
$\chi$. The potential minimum and maximum are located at $\chi = 0$
and $\phi = 0$, respectively.  Here and in what follows we focus on
the real components of $\phi$ and $\chi$ and drop their imaginary
components for simplicity.\footnote{\label{f1} This does not affect the inflation dynamics, but
it may change the efficiency of the preheating, as will be discussed.}  The scalar potential around the maximum
$\phi=0$ can be expressed as
\beq
V(\varphi)\;\simeq\; \mc^2 \lrfp{\mu^2}{\lambda}{\frac{2}{m}} \left(1-\frac{g}{\mu^2} \lrfp{\varphi}{\sqrt{2}}{n} \right)^\frac{2}{m},
\label{vpote}
\eeq
where we have substituted (\ref{susy}) and defined $\varphi \equiv
\sqrt{2} \,{\rm Re}[\phi]$.  The scalar potential around $\varphi = 0$
is so flat that the inflation takes place.  Note that the above
expression (\ref{vpote}) is not valid at $\chi = 0$, because $\phi$
has a large SUSY mass there. The potential near $\chi = 0$ is simply
given by the soft mass term of $\chi$.  See Fig.~\ref{fig:potential}
for the typical shape of the inflaton potential.  The inflation takes
place near the top of the potential, and the inflaton $\phi$ turns
into $\chi$ after inflation because of the SUSY condition given by
(\ref{susy}).  For sufficiently small $|m_\phi^2|$ and $m_\phi^2 < 0$,
the potential minimum and maximum are still located at $\chi = 0$ and
$\phi = 0$ and the inflation is possible along the valley of the
potential.

There are in general other contributions like
\beq
\delta V \;=\; m_{3/2}^2 \lambda \chi^m + m_{3/2}^2 g \phi^n + {\rm h.c.}   \label{VA}
\eeq
which arise from the K\"ahler potential or inserting $\la S \ra =
O(m_{3/2})$ in (\ref{W}).\footnote{ The $S$ is known to develop a VEV
  of order $m_{3/2}$ once the SUSY breaking is taken into
  account~\cite{Dvali:1997uq}.  } However, these terms have negligible
effects on the inflation dynamics as long as $\mc \sim m_{3/2}$.  (We
will see that these terms are important for the AD mechanism later.)

\begin{figure}[t!]
\begin{center}
\includegraphics[scale=0.85]{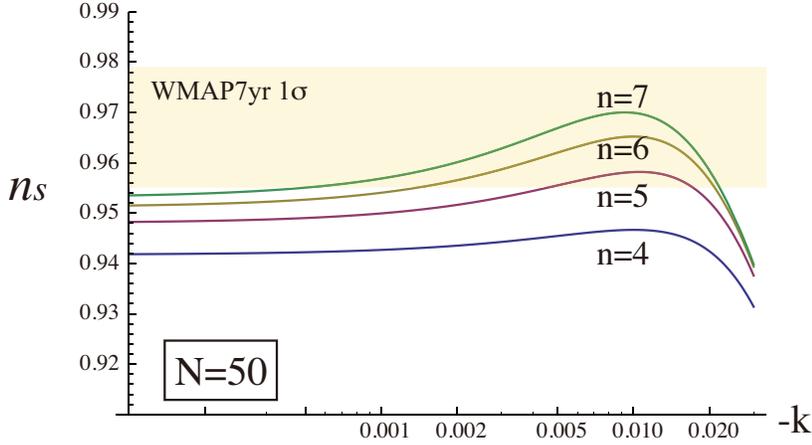}
\caption{The spectral index for $n = 4,5,6$ and $7$.
 The shaded region shows the $1 \sigma$ allowed range,  $n_s = 0.968 \pm 0.012$,
by the WMAP 7yr data~\cite{Komatsu:2010fb}. }
\label{ns}
\end{center}
\end{figure}

Let us study the inflationary dynamics in detail. To simplify our analysis we focus on the inflaton dynamics around the top of the potential,
$\varphi \ll (\mu^2/g)^{1/n}$.
The potential can then be approximated as
\bea
V(\varphi) &\simeq& \mc^2 \lrfp{\mu^2}{\lambda}{\frac{2}{m}}  \left(1- \frac{2g}{m \mu^2} \lrfp{\varphi}{\sqrt{2}}{n}  \right)
+\frac{1}{2}m_\phi^2 \varphi^2,\\
&\equiv& V_0 - \kappa \varphi^n+\frac{1}{2} k V_0 \varphi^2.
\label{vinf}
\eea
For successful inflation, we require $m_\phi^2 < 0$ and $|m_\phi^2|\ll
V_0$, or equivalently, $k<0$ and $|k| \ll 1$.  If this is satisfied,
the inflation takes place at around $\varphi = 0$ and the inflaton
$\varphi$ turns into $\chi$ after inflation and stabilized at $\chi =
0$.  The fine-tuning of $k$ is nothing but the so-called
$\eta$-problem. We allow such fine-tuning if it is required for
successful inflation.  The inflation ends at $\varphi = \varphi_{\rm
  end}$ given by
\beq
\varphi_{\rm end}\;=\; \lrfp{V_0(1+k)}{n(n-1) \kappa}{\frac{1}{n-2}}.
\eeq
The position of the inflaton when the WMAP pivot scale exited the horizon is
\beq
\varphi_N^{n-2} \;=\; \frac{k V_0}{n\kappa}\left[1+\left(\frac{(n-1)k}{1+k}-1\right)e^{-N(n-2)k}\right]^{-1},
\eeq
where $N$ denotes the e-folding number. The scalar spectral index is evaluated as
\beq
	n_s = 1+2k\left[ 
		1-\frac{n-1}{1+\left(\frac{(n-1)k}{1+k}-1\right)e^{-N(n-2)k}}
	\right].
\eeq
In the limit of $|k|\ll 1$, we obtain
\beq 
	n_s \simeq 1-\frac{2(n-1)}{N(n-2)+n-1}.
	\label{nsk=0}
\eeq
The spectral index for $n = 4,5,6$ and $7$ with $N=50$ is shown in
Fig.~\ref{ns}. We can see that the scalar spectral index is consistent
with the WMAP result~\cite{Komatsu:2010fb}, $n_s = 0.968 \pm 0.012$,
for $n \geq 5$, while the predicted value of $n_s$ for $n=4$ has a
slight tension with the WMAP result. This tension can be easily
alleviated by the Coleman-Weinberg correction to the scalar potential
if $\phi$ has a sizable Yukawa coupling~\cite{Nakayama:2012dw}.

The WMAP normalization is given by~\cite{Komatsu:2010fb}
\beq
\frac{1}{12 \pi^2} \frac{V(\varphi)^3}{V'(\varphi)^2} \;\simeq\; 2.43 \times 10^{-9}.  \label{WMAP}
\eeq
For given parameters in the superpotential (\ref{W}), the WMAP
normalization condition (\ref{WMAP}) fixes the soft SUSY breaking
mass, $m_\chi$, and hence the Hubble scale during inflation, $H_{\rm
  inf}$.

\begin{figure}[t]
\begin{center}
\includegraphics[scale=0.6]{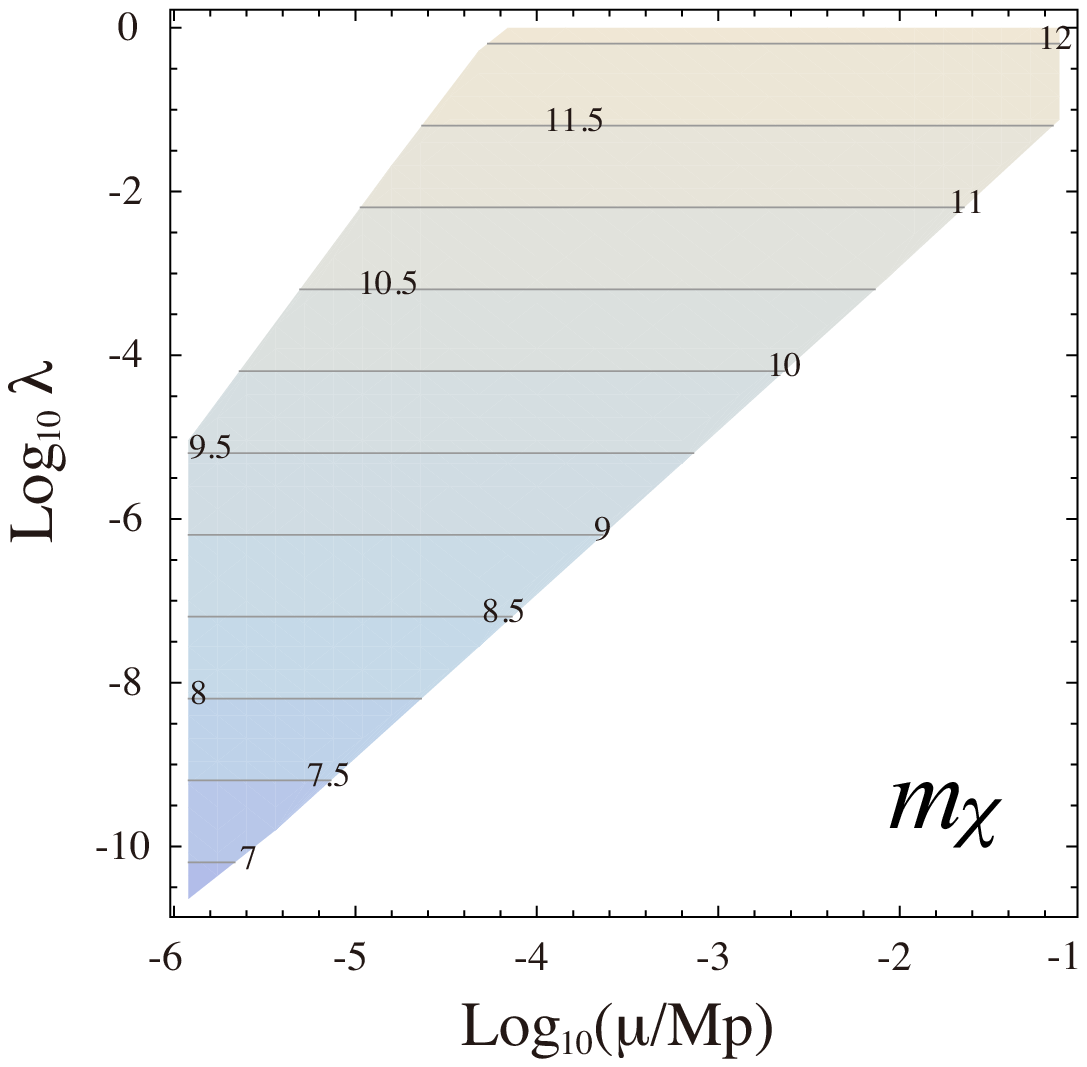}\qquad
\includegraphics[scale=0.6]{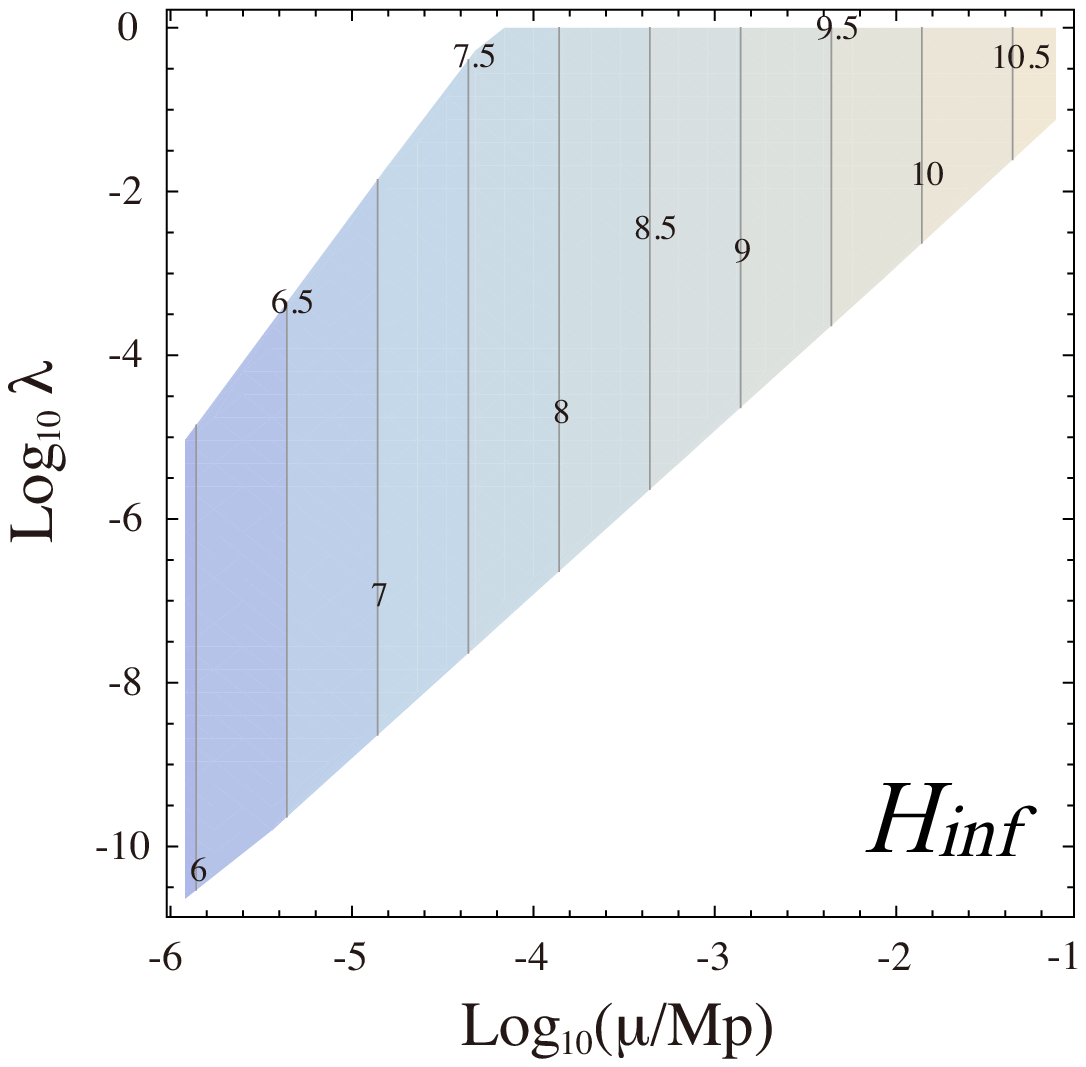}
\caption{The contours of $\log_{10}[\mc/{\rm GeV}]$ (left) and 
$\log_{10}[H_{\rm inf}/{\rm GeV}]$ (right) where we set $g=1$, $m=2$, $n=4$ and $N=50$.}
\label{fig1}
\end{center}
\end{figure}

\begin{figure}[t]
\begin{center}
\includegraphics[scale=0.6]{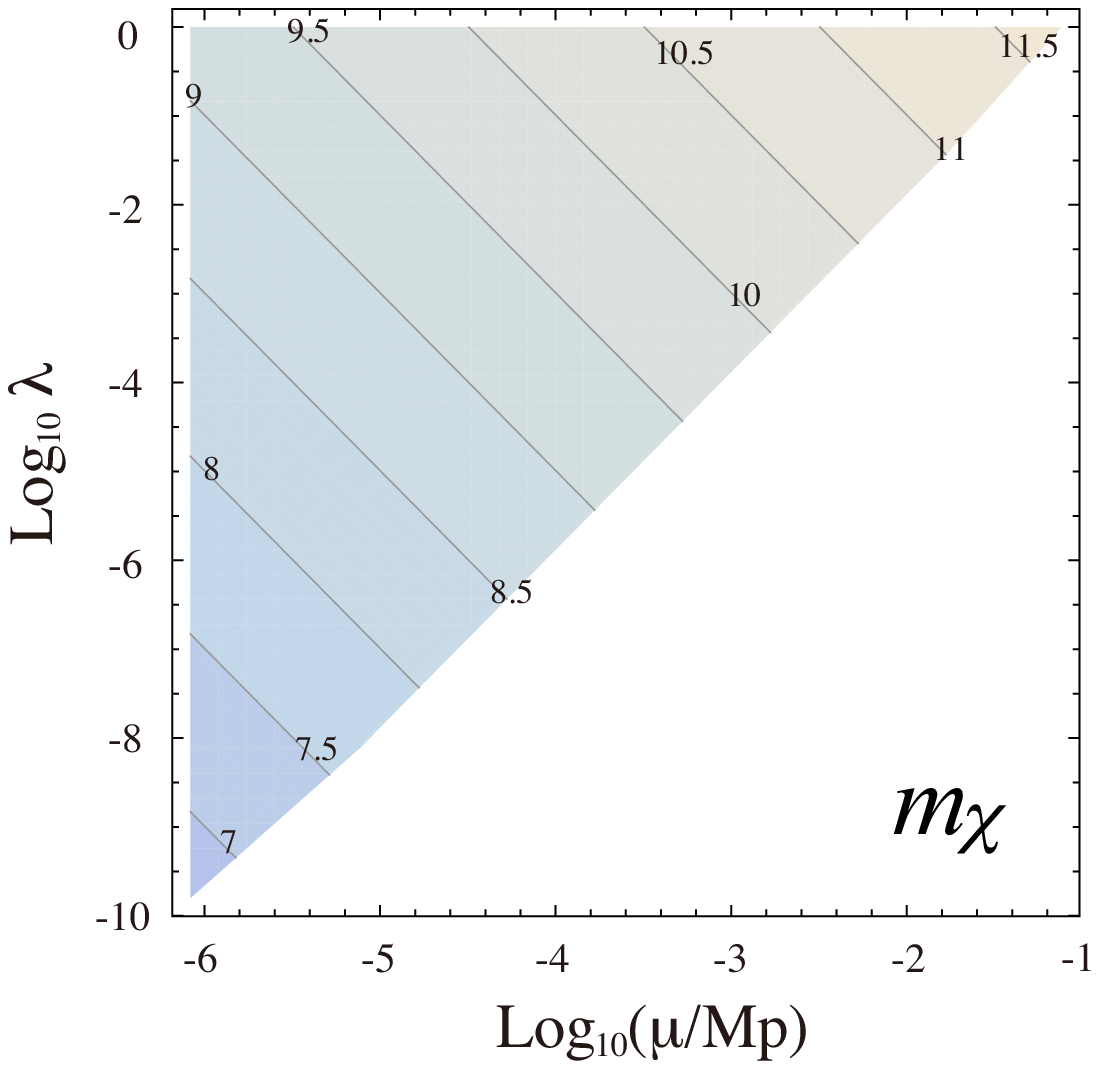} \qquad
\includegraphics[scale=0.6]{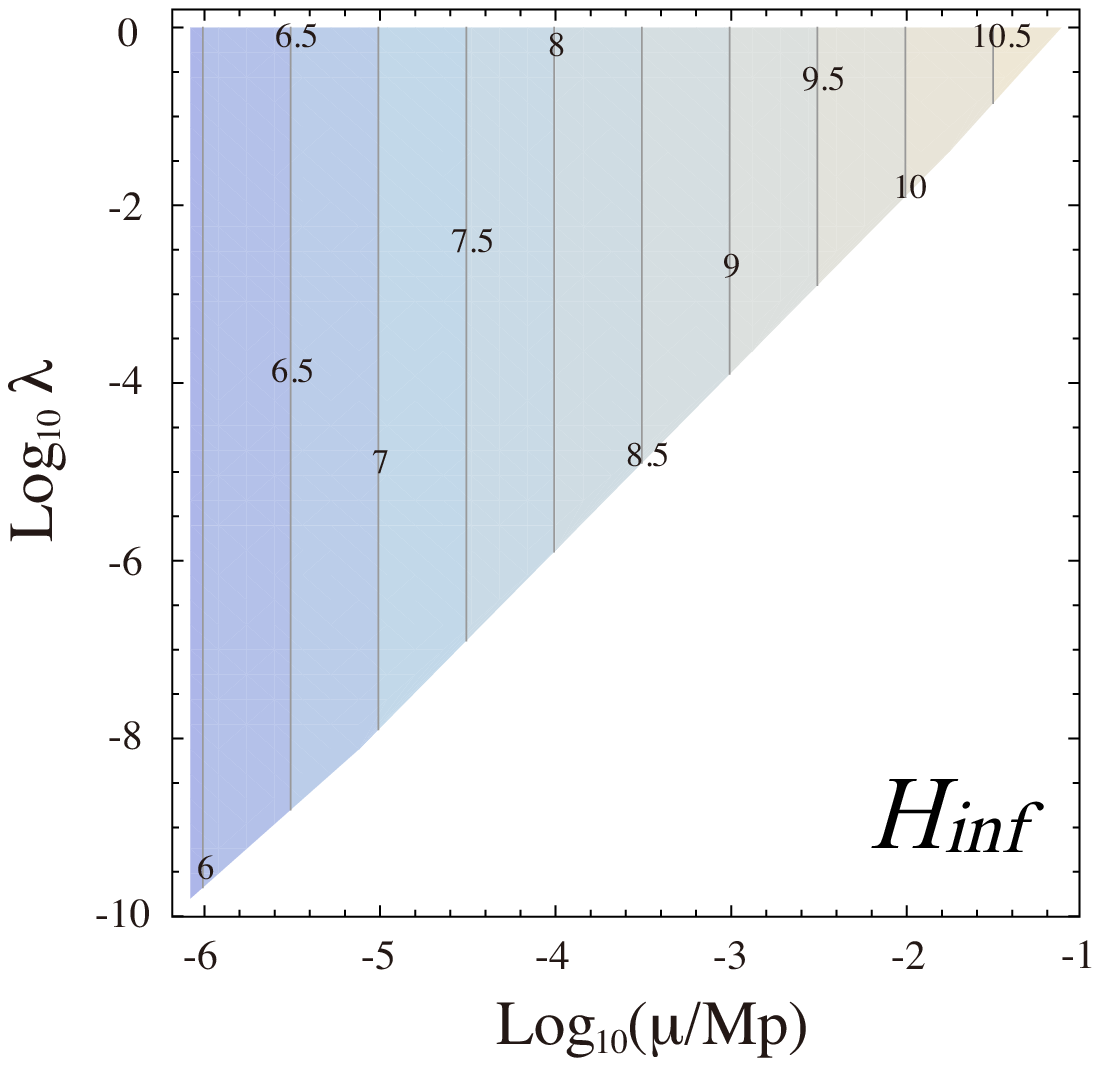}
\caption{Same as Fig.1 but for $m=4$.}
\label{fig2}
\end{center}
\end{figure}

\begin{figure}[t]
\begin{center}
\includegraphics[scale=0.6]{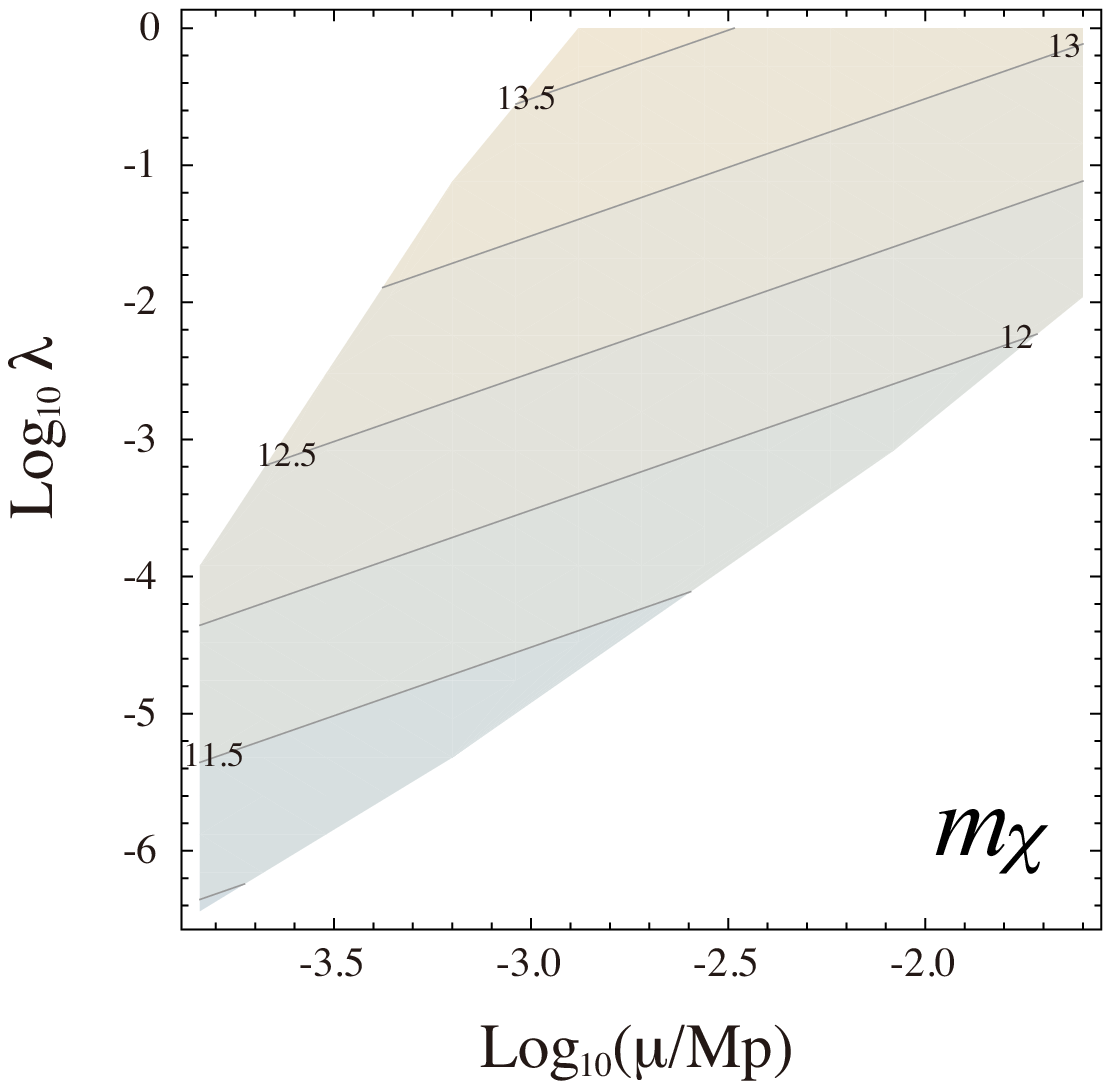} \qquad
\includegraphics[scale=0.6]{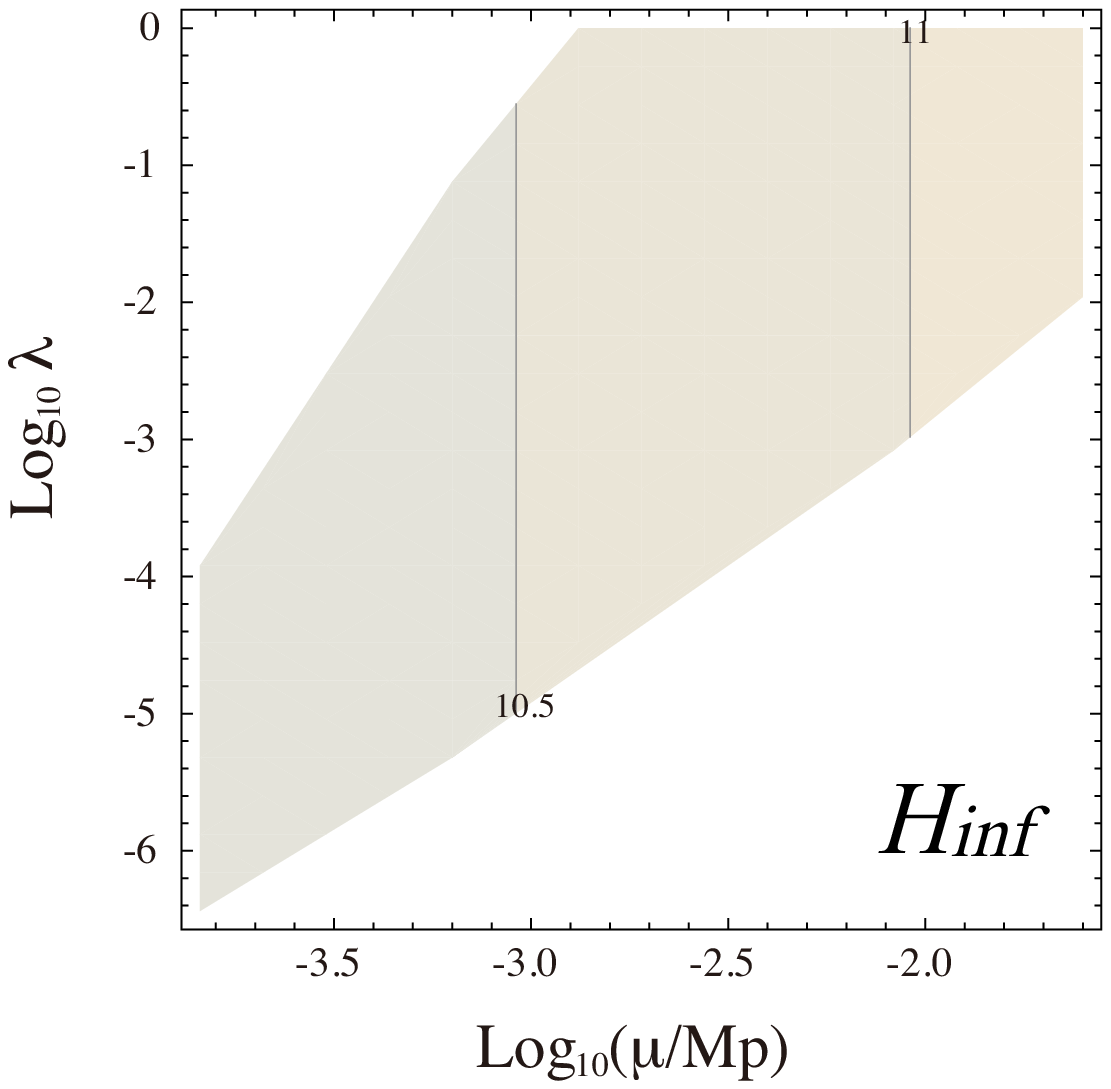}
\caption{Same as Fig.1 but for $m=2$ and $n=6$.}
\label{fig3}
\end{center}
\end{figure}

In Fig.~\ref{fig1} we show the contours of $\log_{10}[\mc/{\rm GeV}]$
and $\log_{10}[H_{\rm inf}/{\rm GeV}]$.  We can see that the soft SUSY
breaking mass scale can be as low as about $\GEV{7}$. Here we have
imposed a couple of constraints: the VEV of $\chi$ and $\phi$ should
be smaller than $0.3 M_p$; the mass orthogonal to the F-flat direction
(\ref{susy}) is at least $3$ times heavier than $\mc$.  Similarly we
show the contours for the case of $m=n=4$ in Fig.~\ref{fig2}. The
range of the soft mass ranges from $\GEV{7}$ to $\GEV{9}$ for $\mu =
\GEV{12}$.   If we take a large $g$,
say, $g = (4 \pi)^2$, the soft mass can be below $\GEV{6}$.  
For $n=6$, the lower bound on the soft mass increases to
$\GEV{11}$ as one can see from Fig.~\ref{fig3}.  It is interesting
that the range of the soft SUSY breaking mass agrees with those for
which $125$\,GeV Higgs mass suggested by the recent ATLAS and CMS
experiments~\cite{ATLAS:2012ae,Chatrchyan:2012tw} can be
realized~\cite{Giudice:2011cg}.

 Note that $H_{\rm inf} < m_\chi (\sim m_{3/2})$ always holds in our
 model as long as $\chi_0 < M_P$.  Therefore, for successful
 inflation, we need a tuning such that $|m_\phi| \ll H_{\rm inf} <
 m_\chi$.  As we can see from Figs.~\ref{fig1}-\ref{fig3}, the
 required amount of tuning is mild around the lower left region, and 
$m_\phi$ must be suppressed by about two orders of magnitude compared
to its natural value.

So far we have not assumed the nature of $\chi$.  An interesting
possibility is that it consists of the D-flat direction of the SSM
fields.  As the simplest case, we can identify it with $\chi^2 = H_u
H_d$.  Then the inflaton transmutes into Higgs after inflation!  This
indicates that the efficient energy transfer to the visible sector
takes place even if the $\phi$ is a gauge singlet and has no sizable
interactions with SSM fields. The process of reheating will be
discussed in the next section.  It is also straightforward to identify
$\chi^m$ with other flat directions such as $(\bar u\bar d\bar d)^2$
and $(LL\bar e)^2$.  In this case, the AD baryogenesis takes place
naturally.

\section{Discussion}
\label{sec:3}

\subsection{Reheating}

An astonishing feature of the present inflation model is that the
inflaton $\phi$ automatically transforms into another field $\chi$
after inflation.  It is $\chi$ rather than $\phi$ that reheats the
Universe.  Therefore, even if $\phi$ is a singlet and its interaction
to the SSM particles is suppressed, the efficient reheating can take
place if $\chi$ has sizable interactions with the SSM particles, or
$\chi$ itself can be the SSM field.  Actually, $\chi$ can be
identified with one of the D-flat directions in the SSM, like $\chi^2
= H_u H_d$ or $\chi^3 = LL\bar e$, etc.  Here we show that the
reheating is efficient in such a case because $\chi$ oscillates around
the enhanced symmetry point, $\chi=0$, and efficient particle
production takes place in each oscillation.  This should be contrasted
to the conventional new inflation models, in which the inflaton
oscillates around the large VEV and its perturbative decay rate can be
suppressed.\footnote{ The tachyonic preheating genetically takes place
  at the end of new inflation, however, the reheating is usually
  competed by the perturbative decay of the inflaton particle.  }

Let us collectively denote the fields that couple to $\chi$ by $\psi$.
The $\psi$ can be SSM (s)fermions and gauge bosons depending on the
properties of $\chi$.  Hereafter, we consider $\psi$ which has the
strongest couplings to $\chi$ and assume the couplings of the
following form,
\bea
\mathcal L &=& \left\{
\bear{cl}
y^2 |\chi|^2 |\tilde q|^2 &~ {\rm for\, \,a\,\,sfermion~}(\psi = \tilde q) \\
y \chi q \bar q + {\rm h.c.} &~ {\rm for\, \,a\,\,fermion~}(\psi = q) \\
y^2 |\chi|^2 A_\mu A^\mu &~ {\rm for\, \,a\,\,gauge\,\,boson~}(\psi = A_\mu) \\
\eear
\right.
\eea
The mass of $\psi$ is $\chi$-dependent, and is given by $m_\psi \sim y
\la \chi \ra$.  Following discussion does not depend on whether $\psi$
is a boson or fermion.  For example, if $\chi$ consists of the $H_u
H_d$, $\psi$ represents top (s)quarks or weak gauge bosons, and the
coupling $y$ denotes either top Yukawa coupling or SU(2) gauge
coupling.  The exact form of the interactions does not matter in the
following discussion.

After inflation ends, $\chi$ oscillates around the minimum $\chi=0$ as
$\chi(t)=\chi_0\sin (m_\chi t )$ (here $t=0$ is set to be the instant
when $\chi$ reaches to minimum $\chi=0$).  We neglect the cosmic
expansion because $m_\chi \gg H_{\rm inf}$ and because we are only
interested in the dynamics within the time scale of dozens of $\chi$
oscillations.  The $\psi$ field obtains a large mass through the
coupling with $\chi$ and the perturbative decay of $\chi$ into $\psi$
is prohibited during most of the time of oscillations except for the
vicinity of $\chi\sim 0$, where the adiabaticity of the $\psi$ is
violated~\cite{Kofman:1997yn}.  The frequency of the $\psi$-particle
with a wavenumber $k$ is given by $\omega_k =\sqrt{y^2\chi^2+k^2}$.
(This $k$ should not be confused with the one in the inflaton
potential (\ref{vinf}).)  As $\chi$ passes the origin, the
adiabaticity is violated, i.e., $|\dot \omega_k / \omega_k^2| > 1$,
and the $\psi$-particle is produced.  The typical wavenumber $k_*$ of
the produced $\psi$-particle is estimated as $k_* =
\sqrt{y\chi_0m_\chi}$.  Therefore, each time the $\chi$ passes through
the origin, the number density of $\psi$ increases by
\begin{equation}
	n_\psi \simeq \frac{k_*^3}{8\pi^3} \simeq \frac{(y\chi_0 m_\chi)^{3/2}}{8\pi^3}.
\end{equation}
The produced $\psi$ particle becomes massive as $\chi$ goes far away
from the origin, and then the $\psi$ decays into light states. The
decay rate of $\psi$ is estimated as $\Gamma_\psi \sim
(f^2/8\pi)m_\psi$, where $f$ is either a gauge or Yukawa coupling of
$\psi$, and $m_\psi \simeq y\chi$.  Thus $\psi$ decays at $t=t_{\rm
  dec} \sim \sqrt{8\pi/(f^2y \chi_0m_\chi)}$ at which it has a mass of
$m_\psi \sim \sqrt{8\pi y \chi_0m_\chi}/f$.\footnote{ Note that
  $t_{\rm dec} > 1/k_*$ and hence the implicit assumption that the
  $\psi$ decays after the particle production ends is justified.  }
This happens much before the $\chi$ returns again back to $\chi=0$.
Therefore, in each oscillation, the energy of $\chi$ decreases by
\beq
	\frac{\Delta \rho_\chi}{\rho_\chi} \simeq \frac{m_\psi(t_{\rm dec}) n_\psi}{m_\chi^2 \chi_0^2/2} 
	\sim \frac{y^2}{\sqrt{2}\pi^{5/2}f}.
\eeq
This is $O(0.1-0.01)$ depending on the couplings of $\chi$ and $\psi$.
Therefore, the $\chi$ energy density is efficiently transferred to the
radiation within dozens of oscillations through the $\psi$ particle
production and its subsequent decay.  The whole process is what is
called the instant preheating~\cite{Felder:1998vq}.  (See also
Ref.~\cite{Allahverdi:2011aj}.)  
If the reheating is completed by the instant preheating, 
the reheating temperature is  estimated as $T_R \sim \sqrt{H_{\rm inf}}$, and it is given by
$T_R\sim 10^{12}$\,GeV for $H_{\rm inf} \sim 10^6$\,GeV.  Thermal
leptogenesis works for such a high reheating
temperature~\cite{Fukugita:1986hr}. 
Note that, as the radiation energy increases 
through the instant preheating, the $\psi$-particle acquires a thermal mass,
which can make the energy transfer inefficient and therefore can terminate the preheating
process  when the energy density of $\chi$ becomes comparable to that of
radiation.  If such back reaction is relevant,  the reheating is completed by the usual perturbative decay. 
In this case, the reheating temperature is 
considered to be about the soft SUSY breaking mass, $T_R \sim m_\chi$. 

So far we have assumed that $\chi$ passes the origin. This may not be the case
if $\chi$ acquires an angular momentum in its complex plane. In fact, if we take
account of the imaginary component of $\phi$, it acquires a certain angular momentum
during and after inflation, which is transferred to $\chi$ through the SUSY condition (\ref{susy}). 
In this case, the preheating process may become inefficient and the perturbative decay
will complete the reheating. 

 Note that SUSY particles can be produced both thermally and
 non-perturbatively.  Since the SUSY breaking scale, which is
 characterized by $m_{3/2}$, is heavier than PeV
 in this model, the abundance of the lightest SUSY particle (LSP)
 likely exceeds the dark matter abundance if the R-parity is
 conserved. The LSP overabundance can be avoided if the R-parity is
 violated by a small amount so that the LSP decays before the big-bang
 nucleosynthesis (BBN).  If the reheating temperature is higher than $m_{3/2}$, 
 gravitinos are also efficiently produced by
 thermal scatterings, and they decay into LSPs much before the BBN and
 those LSPs quickly decay through the R-parity violating
 interactions. Alternatively, the cosmic density of the LSP can be
 suppressed if the LSP mass is sufficiently light as in the case of the axino LSP.

\subsection{Initial condition}

In order for our inflation scenario to work, the initial position of
the inflaton must be near the maximum $\phi=0$ along the flat
direction (\ref{susy}).  This initial condition is dynamically
realized by considering the pre-inflation which occurs in the same
superpotential (\ref{W}).  First note that the scalar potential is
flat along $S$ for $\phi=\chi= 0$ where the potential energy is given
by $\mu^4$.  Thus the pre-inflation takes place for $\phi\sim\chi\sim
0$ with a sufficiently large value of $S$, and $S$ plays the role of
inflaton.  We assume $\phi$ is stabilized at the origin by a positive
Hubble-induced mass during the pre-inflation.\footnote{ Even if $\phi$
  is not stabilized at the origin during pre-inflation, it may settle
  down at the origin after the pre-inflation through its interactions
  with the decay product of $S$ or $\chi$. } If $m=2$, $\chi$ is
stabilized at the origin until $\chi$ becomes tachyonic when $S \sim
\mu/\sqrt{\lambda}$ and the pre-inflation ends. The inflationary
dynamics is same as the hybrid inflation~\cite{Copeland:1994vg}.  If
$m > 2$, $\chi$ develops a small but non-zero $S$-dependent VEV, and
the inflation ends when $S$ becomes of order $
(\mu^2/\lambda)^{1/(2m-2)}$.  as in the smooth hybrid inflation
model~\cite{Lazarides:1995vr}. In both cases, it is $\chi$ that
develops large VEV at the end of pre-inflation while $\phi=0$.  Thus
the desired initial condition for the new inflation is dynamically
achieved.

\subsection{Extensions to other D-flat directions, and AD baryogenesis}

As already noticed, the $\chi$ field can be identified with other
D-flat directions in the SSM, and if it has non-zero baryon and/or
lepton numbers, it can create the baryon asymmetry through the AD
mechanism.  

As we have mentioned in Footnote~\ref{f1}, $\chi$ can acquire a non-zero
angular momentum from the inflaton dynamics. This however depends on the
initial condition of the imaginary component of the inflaton $\phi$. Let us here
concentrate on the dynamics of $\chi$ after inflation.
 For example, we consider $\chi^6 = (LL\bar e)^2$.  The
AD field obtains a scalar potential of the form given by (\ref{VA}),
which serves as the source of the baryon number violation and induces
an angular motion on the complex plane of $\chi$.  Although the AD
field receives many angular kicks because of $m_\chi \gg H_{\rm inf}$,
the first one gives the dominant contribution. The resultant baryon
asymmetry is estimated as
\beq
\frac{n_B}{s}\;\sim\; \frac{T_R}{H_{\rm inf}^2} \frac{ \lambda m_{3/2}^2 \chi_0^m}{H_{\rm inf}} \delta
\eeq
where we have defined $\chi_0 \equiv (\mu^2/\lambda)^{1/m}$, and
$\delta (<1)$ represents the CP phase. The particle production becomes
inefficient if the $\chi$ has large angular momentum, because the
$\chi$ does not pass the origin.  If the non-perturbative decay of
$\chi$ is suppressed, the reheating temperature is considered to be of
order $m_\chi$. Therefore $T_R$ should be in the range between
$m_\chi$ and $\sqrt{H_{\rm inf}}$. We have found that in the case of
$m=6$, $n=4$ and $\delta \sim O(0.1)$, the above baryon asymmetry
varies from $O(10^{-9})$ to $O(10^{-3})$ in the allowed region of
$\mu$ and $\lambda$.  Therefore, it is possible to generate a right
amount of the baryon asymmetry via the AD mechanism.  The magnitude of
the baryon isocurvature perturbation is given by $S_b \sim H_{\rm
  inf}/(2 \pi \chi_0)$ and it is much smaller than the observational
constraint. We have confirmed it is smaller than $10^{-8}$ for the
above case.

\subsection{Other applications}
Let us comment on other applications of our model.  One extension of
the model is consider an exponential dependence of $\phi$.
\bea
	W &=& S(\mu^2 - \lambda \chi^m -ge^{-b\phi}).
\eea
The K\"ahler potential is considered to respect a shift symmetry of
$\phi$: $\phi \rightarrow \phi+ i \alpha$, where $\alpha$ is a real
transformation parameter. The inflationary dynamics is similar to the
case studied in Sec.~\ref{sec:2}, and some results can be obtained
simply by taking a limit of $n \rightarrow \infty$.  (For instance,
$n_s$ can be obtained by taking this limit in Eq.~(\ref{nsk=0}).)
Interestingly, if we take $m=2$ and $\chi^2 = H_u H_d$, the resultant
potential is quite similar to the Higgs
inflation~\cite{Bezrukov:2007ep}. The great difference is that,
although the potential is similar, there is no large coupling in our
model, avoiding the issue of the unitarity.

It is also possible to use the moduli space (\ref{susy}) as a
curvaton~\cite{Lyth:2001nq}.  In particular, the so-called hilltop
curvaton can be easily realized because the potential maximum is a
symmetry-enhanced point~\cite{Kawasaki:2008mc,Kawasaki:2011pd}.

So far we have focused on the case of $n\geq 3$.  In the case of
$n=2$, the sufficient amount of inflation does not occur, but thermal
inflation may take place, if $\phi$ acquires a thermal mass through
its interactions with the ambient plasma.\footnote{ Similar arguments
  have been made in \cite{Hindmarsh:2012wh} in the context of
  anomaly-mediation with U(1) extension to solve the tachyonic slepton
  problem.  } The reheating is induced by the decay of $\chi$ in this
case, and the resultant reheating temperature will be so high that the
leptogenesis will be possible.  The reheating process is similar to
that described before.

\section{Conclusions}
In this letter, we have proposed a new inflation model, in which the
SM gauge singlet inflaton  turns into the Higgs field (or more
precisely, $H_u H_d$) after inflation. The energy stored in the
inflaton is directly transferred to the $H_u H_d$ condensate.  This is
a novel type of reheating; instead of producing the SM particles from
the inflaton decay, the inflaton energy is directly transmuted into
the $H_u H_d$ flat direction.  The transmutation is due to the
inflaton dynamics in the moduli space of supersymmetric vacua spanned
by the inflaton and the Higgs field.  Interestingly, because of this
transmutation, the new inflation ends up in a symmetry vacuum of the
Higgs boson, which enables rapid thermalization of the Higgs
condensate.  The initial condition for the successful inflation can be
naturally realized by the pre-inflation in which the Higgs field plays
the role of a waterfall field. It is possible to replace the role of $H_uH_d$ with
other D-flat directions in SSM. In particular, we have also pointed out that the 
AD baryogenesis naturally takes place in this case.

\section*{Acknowledgment}
 This work was supported by the Grant-in-Aid for Scientific Research
 on Innovative Areas (No.24111702 [FT], No.21111006 [KN and FT] and
 No.23104008 [FT]), Scientific Research (A) (No.22244030 [KN and FT]
 and No.21244033 [FT]), and JSPS Grant-in-Aid for Young Scientists (B)
 (No.24740135) [FT].  This work was also supported by World Premier
 International Center Initiative (WPI Program), MEXT, Japan.


\begin{thebibliography}{99}

\bibitem{Guth:1980zm}
  A.~H.~Guth,
  Phys.\ Rev.\  {\bf D23}, 347-356 (1981);
A.~A.~Starobinsky,
Phys.\ Lett.\ B {\bf 91} (1980) 99;
  K.~Sato,
  Mon.\ Not.\ Roy.\ Astron.\ Soc.\  {\bf 195}, 467-479 (1981).

\bibitem{Komatsu:2010fb}
E.~Komatsu {\it et al.} [WMAP Collaboration],
Astrophys.\ J.\ Suppl.\ {\bf 192} (2011) 18
[arXiv:1001.4538 [astro-ph.CO]].

  \bibitem{Weinberg:zq}
    S.~Weinberg,
    Phys.\ Rev.\ Lett.\  {\bf 48}, 1303 (1982).
    
\bibitem{Krauss:1983ik}
    L.~M.~Krauss,
    Nucl.\ Phys.\ B {\bf 227}, 556 (1983).
    
    
\bibitem{Endo:2006qk} 
  M.~Endo, M.~Kawasaki, F.~Takahashi and T.~T.~Yanagida,
  Phys.\ Lett.\ B {\bf 642}, 518 (2006)
  [hep-ph/0607170].
  
  
\bibitem{Endo:2007ih} 
  M.~Endo, F.~Takahashi and T.~T.~Yanagida,
  Phys.\ Lett.\ B {\bf 658}, 236 (2008)
  [hep-ph/0701042].
  
\bibitem{Kawasaki:2006gs} 
  M.~Kawasaki, F.~Takahashi and T.~T.~Yanagida,
  Phys.\ Lett.\ B {\bf 638}, 8 (2006)
  [hep-ph/0603265];
  Phys.\ Rev.\ D {\bf 74}, 043519 (2006)
  [hep-ph/0605297].
  
\bibitem{Asaka:2006bv} 
  T.~Asaka, S.~Nakamura and M.~Yamaguchi,
  Phys.\ Rev.\ D {\bf 74}, 023520 (2006)
  [hep-ph/0604132].
  
\bibitem{Endo:2007sz} 
  M.~Endo, F.~Takahashi and T.~T.~Yanagida,
  Phys.\ Rev.\ D {\bf 76}, 083509 (2007)
  [arXiv:0706.0986 [hep-ph]].
  
\bibitem{Kallosh:2004yh}
  R.~Kallosh, A.~D.~Linde,
  JHEP {\bf 0412}, 004 (2004).
  [hep-th/0411011].

\bibitem{Copeland:1994vg} 
  E.~J.~Copeland, A.~R.~Liddle, D.~H.~Lyth, E.~D.~Stewart and D.~Wands,
  Phys.\ Rev.\ D {\bf 49}, 6410 (1994)
  [astro-ph/9401011];
  G.~R.~Dvali, Q.~Shafi and R.~K.~Schaefer,
  Phys.\ Rev.\ Lett.\  {\bf 73}, 1886 (1994)
  [hep-ph/9406319].
  

\bibitem{Lazarides:1995vr} 
  G.~Lazarides and C.~Panagiotakopoulos,
  Phys.\ Rev.\ D {\bf 52}, 559 (1995)
  [hep-ph/9506325].
  
  

\bibitem{Affleck:1984fy} 
  I.~Affleck and M.~Dine,
  Nucl.\ Phys.\ B {\bf 249}, 361 (1985).
  
\bibitem{Dine:1995uk}
  M.~Dine, L.~Randall, S.~D.~Thomas,
  Phys.\ Rev.\ Lett.\  {\bf 75 }  398 (1995) 
  [arXiv:hep-ph/9503303]; 
  M.~Dine, L.~Randall and S.~D.~Thomas,
  Nucl.\ Phys.\  B {\bf 458}, 291 (1996)
  [arXiv:hep-ph/9507453].


\bibitem{Allahverdi:2006iq} 
  R.~Allahverdi, K.~Enqvist, J.~Garcia-Bellido and A.~Mazumdar,
  Phys.\ Rev.\ Lett.\  {\bf 97}, 191304 (2006)
  [hep-ph/0605035];
  R.~Allahverdi, K.~Enqvist, J.~Garcia-Bellido, A.~Jokinen and A.~Mazumdar,
  JCAP {\bf 0706}, 019 (2007)
  [hep-ph/0610134].
  
\bibitem{Linde:1981mu}
  A.~D.~Linde,
  Phys.\ Lett.\ B {\bf 108}, 389 (1982);


\bibitem{Dvali:1997uq} 
  G.~R.~Dvali, G.~Lazarides and Q.~Shafi,
  Phys.\ Lett.\ B {\bf 424}, 259 (1998)
  [hep-ph/9710314].



\bibitem{Nakayama:2012dw}
K.~Nakayama and F.~Takahashi,
JCAP 05(2012)035
[arXiv:1203.0323 [hep-ph]];
see also 
JCAP {\bf 1110} (2011) 033
[arXiv:1108.0070 [hep-ph]];
Phys.\ Lett.\ B {\bf 707} (2012) 142
[arXiv:1108.3762 [hep-ph]].





\bibitem{ATLAS:2012ae}
  G.~Aad {\it et al.}  [ATLAS Collaboration],
  Phys.\ Lett.\  B {\bf 710}, 49 (2012)
  [arXiv:1202.1408 [hep-ex]].
  
\bibitem{Chatrchyan:2012tw}
  S.~Chatrchyan {\it et al.}  [CMS Collaboration],
  Phys.\ Lett.\  B {\bf 710}, 403 (2012)
  [arXiv:1202.1487 [hep-ex]].
  

\bibitem{Giudice:2011cg} 
  G.~F.~Giudice and A.~Strumia,
  Nucl.\ Phys.\ B {\bf 858}, 63 (2012)
  [arXiv:1108.6077 [hep-ph]];
  G.~Degrassi, S.~Di Vita, J.~Elias-Miro, J.~R.~Espinosa, G.~F.~Giudice, G.~Isidori and A.~Strumia,
  arXiv:1205.6497 [hep-ph].

    
\bibitem{Kofman:1997yn} 
  L.~Kofman, A.~D.~Linde and A.~A.~Starobinsky,
  Phys.\ Rev.\ D {\bf 56}, 3258 (1997)
  [hep-ph/9704452].

\bibitem{Felder:1998vq} 
  G.~N.~Felder, L.~Kofman and A.~D.~Linde,
  Phys.\ Rev.\ D {\bf 59}, 123523 (1999)
  [hep-ph/9812289].
  
\bibitem{Allahverdi:2011aj} 
  R.~Allahverdi, A.~Ferrantelli, J.~Garcia-Bellido and A.~Mazumdar,
  Phys.\ Rev.\ D {\bf 83}, 123507 (2011)
  [arXiv:1103.2123 [hep-ph]].
  
\bibitem{Fukugita:1986hr} 
  M.~Fukugita and T.~Yanagida,
  Phys.\ Lett.\ B {\bf 174}, 45 (1986).
  
  
\bibitem{Lyth:2001nq}
  D.~H.~Lyth and D.~Wands,
  Phys.\ Lett.\  B {\bf 524}, 5 (2002)
  [arXiv:hep-ph/0110002];
  T.~Moroi and T.~Takahashi,
  Phys.\ Lett.\  B {\bf 522}, 215 (2001)
  [Erratum-ibid.\  B {\bf 539}, 303 (2002)]
  [arXiv:hep-ph/0110096];
  K.~Enqvist and M.~S.~Sloth,
  Nucl.\ Phys.\  B {\bf 626}, 395 (2002)
  [arXiv:hep-ph/0109214].
 
\bibitem{Bezrukov:2007ep}
  F.~L.~Bezrukov, M.~Shaposhnikov,
  Phys.\ Lett.\  {\bf B659}, 703-706 (2008).
  [arXiv:0710.3755 [hep-th]]. 
  
\bibitem{Kawasaki:2008mc}
M.~Kawasaki, K.~Nakayama and F.~Takahashi,
JCAP {\bf 0901} (2009) 026
[arXiv:0810.1585 [hep-ph]].

\bibitem{Kawasaki:2011pd}
M.~Kawasaki, T.~Kobayashi and F.~Takahashi,
Phys.\ Rev.\ D {\bf 84} (2011) 123506
[arXiv:1107.6011 [astro-ph.CO]].

\bibitem{Hindmarsh:2012wh} 
  M.~Hindmarsh and D.~R.~T.~Jones,
  arXiv:1203.6838 [hep-ph].


  
\end{thebibliography}
\end{document}